\documentclass[aps,prl,twocolumn,showpacs,10pt]{revtex4-1}
\usepackage{graphicx}
\usepackage{amssymb}
\usepackage{amsmath,color,mathrsfs}
\usepackage{bbold}
\usepackage{comment}
\usepackage[dvipsnames]{xcolor}
\usepackage{bm}
\pdfoutput=1
\usepackage{hyperref}
\hypersetup{urlcolor=blue, colorlinks=true, citecolor=blue, linkcolor=blue}
\usepackage{todonotes}

\newcommand{\mysection}[1]{\textit{#1}.---}

\begin{document}

	\title{Lifetime of Majorana qubits in Rashba nanowires with  non-uniform chemical potential}
	\author{Pavel P. Aseev}
	\author{Jelena Klinovaja}
	\author{Daniel Loss}
	\affiliation{Department of Physics, University of Basel, Klingelbergstrasse 82, CH-4056 Basel, Switzerland}
	\date{\today}
	
\begin{abstract}		
We study the lifetime of topological qubits based on Majorana bound states hosted in a one-dimensional Rashba nanowire (NW) with proximity-induced superconductivity
and non-uniform chemical potential needed for manipulation and read-out.
If nearby gates tune the chemical potential locally so that part of the NW is in the trivial phase, Andreev bound states (ABSs) can emerge which are localized at the interface between topological and trivial phases with energies significantly less than the gap. The emergence of such subgap states 
strongly decreases the Majorana qubit lifetime at finite temperatures due to local perturbations that can excite the system into these ABSs. Using Keldysh formalism, we study such excitations  caused by fluctuating charges in capacitively coupled gates and calculate the corresponding Majorana lifetimes  due to thermal noise, which are shown to be much shorter than those in NWs with uniform chemical potential.  
\end{abstract}
	
	\maketitle
	
	\mysection{Introduction} 
The main benefit of topological quantum computing~\cite{Kitaev2003, Kitaev2001, Wilczek2009, Stern, Nayak2008} is the  possibility to encode quantum information in degenerate many-body ground states (GSs) in such a way that it is free of decoherence in the ideal case. A realization of such topologically protected GSs 
are zero-energy Majorana bound states (MBSs) in a topological superconductor (TSC)~\cite{Schnyder2008, Sato2009, Qi2011, Tanaka2012, Vijay2015, Vijay2016}. 
One of the most promising systems are semiconducting Rashba nanowires (NWs) in proximity with an
$s$-wave superconductor and in the presence of magnetic fields~\cite{Alicea2010, Lutchyn2010, Oreg2010, Mourik2012,  Das2012, Deng2012, Sticlet2012, Churchill2013, Rokhinson2012,  Klinovaja2012, Chevallier2012, San-Jose2012, Dominguez2012, Terhal2012, Klinovaja2012a, Prada2012, DeGottardi2013, Thakurathi2013, Maier2014,  Escribano2017, Prada2017, Ptok2017, Kobialka2018, DeMoor2018}. 

In  realistic cases where the system is subject to random state fluctuations, the TSC may be driven out of its GS. This happens if the TSC is coupled to ungapped~\cite{Budich2012} or gapped~\cite{Goldstein2011, Rainis2012} fermionic baths as well as to fluctuating bosonic fields  \cite{Goldstein2011} (\textit{e.g.} phonons  \cite{Goldstein2011, Knapp2018} or electromagnetic environments~\cite{Knapp2018}).
A similar mechanism is due to thermal fluctuations of a gate potential~\cite{Schmidt2012, Lai2018}. In this case the TSC stays in the GS only for a finite lifetime $\tau$ after which 
the quantum state of the qubit associated with MBSs will leak out of the computational
space, resulting in losing the quantum information. The corresponding rate of excitations caused by thermal fluctuations is exponentially small at low temperatures~\cite{Kaplan1976}. Estimations in Ref.~\onlinecite{Schmidt2012} show that in order to maintain $\tau$ as high as microseconds the temperature should not exceed $\Delta/5$, where $\Delta$ is the gap in the TSC (we set $k_B=1$). This estimation is justified for the simple case of uniform system parameters. However, in many schemes of topological quantum computing  the localized MBSs must be braided around each other, which requires to change the parameters locally. For example, in Ref.~\onlinecite{Alicea2011},
the gates control local chemical potentials so that the NW is separated into topological and trivial sections with MBSs at the interfaces, which allows one to manipulate the  non-local fermions  which  form the Majorana qubit.
Recent studies~\cite{Flensberg, Liu2013, Ricco2016, Plugge2016,  Karzig2017, Guessi2017, Xu2017,Hoffman2017} were focused on the case where detection of MBSs and reading out associated qubits can be achieved by tuning gates to couple MBSs to quantum dots~\cite{Golub2011, Zocher2013, Gong2014, Leijnse2014, Vernek2014, Deng2016, Szombati2016, Hoffman2017, Chevallier2018}. Since gates are located near the topological NWs they create a non-uniform electrostatic potential and can tune the ends of the NWs out of the topological phase.

\begin{figure}[t]
		\includegraphics[width=\columnwidth]{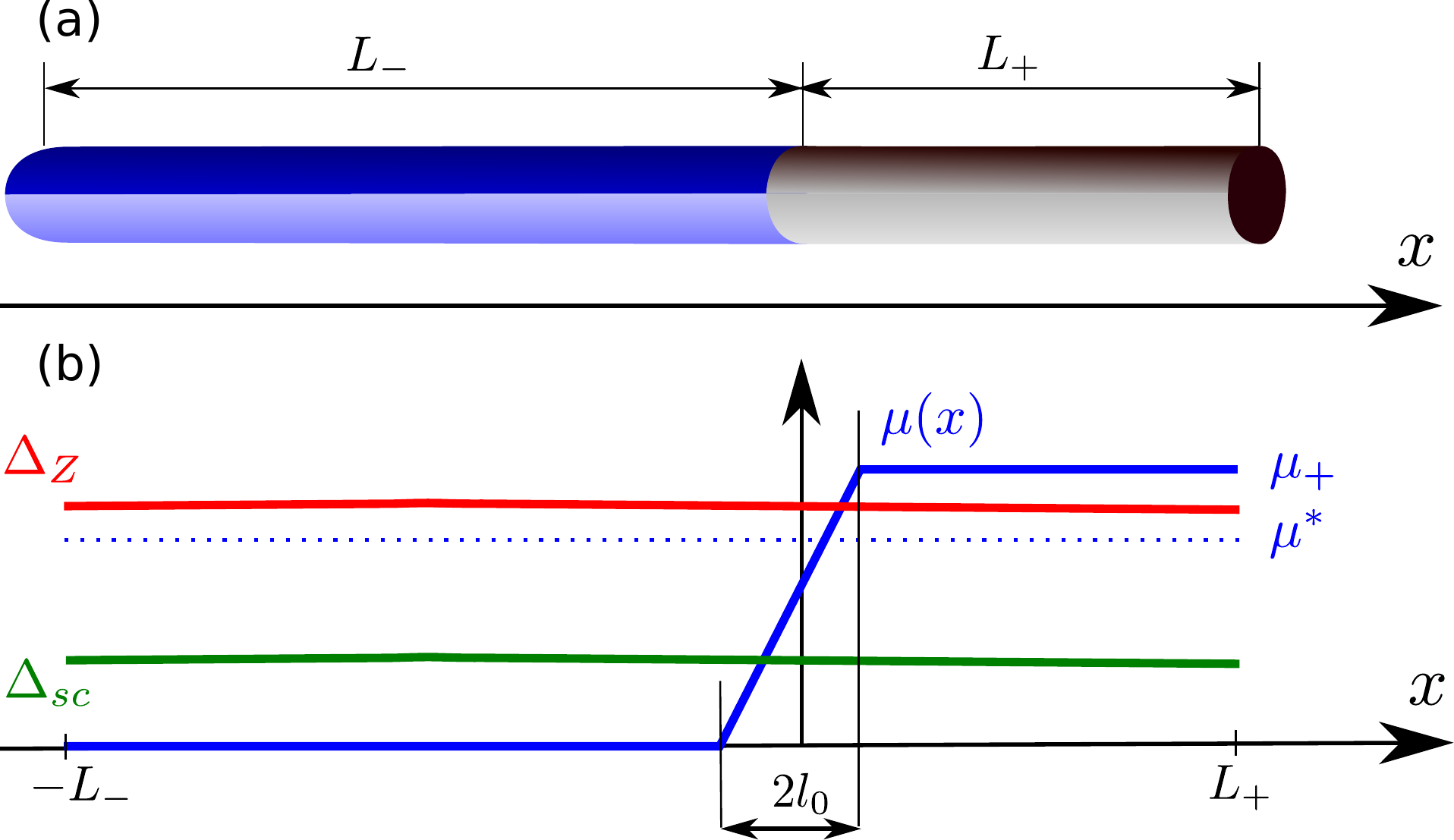}		
\caption{(a) Sketch of  proximitized Rashba NW consisting of two sections. (b) The non-uniform chemical potential $\mu(x)$  is controlled by a nearby gate, so that the right section (gray) of length $L_{+}$ is in the trivial ($|\mu_+|>\mu^*$) or topological ($|\mu_+|<\mu^*$) phase, with $\mu^*$ being the critical value (see text),
while the left section (blue) of length $L_-$ with $\mu=0$ stays always topological. At the interface, $|x|<l_0$, $\mu(x)$ grows linearly in $x$. The proximity  gap $\Delta_{sc}$ and Zeeman energy $\Delta_Z$ are uniform.
		}
		\label{fig:setup}
	\end{figure}

It has been shown in Ref.~\onlinecite{Pedrocchi2015} that  non-localities introduced by braiding restricts the lifetime of qubits due to errors occurring during their motion. In this paper we will show that at finite temperatures the non-uniformity of system parameters can strongly reduce the lifetime  even for immobile MBSs. As a model  we consider a 1D proximitized Rashba NW with a non-uniform chemical potential induced by gates~(see Fig.~\ref{fig:setup}). 
We also study how $\tau$ changes if a part of the NW is tuned out of the topological phase.	  
We develop a microscopic formalism with Keldysh techniques to treat the noisy gates and to calculate $\tau$. This approach can be easily generalized to
other bosonic modes such as plasmons, phonons \textit{etc}. 
Using realistic parameters, we are able to predict the optimal regime of operation for 
Majorana qubits.

\mysection{Model}
We consider a setup consisting of a spinful single-band NW with a proximity-induced superconducting gap $\Delta_{sc}$. The Rashba spin-orbit interaction (SOI) of  strength $\alpha_R$ sets the spin quantization axis to be perpendicular to the NW and corresponds to the SOI energy $E_{so} = m_0\alpha_R^2/(2\hbar^2)$, where $m_0$ is the effective electron mass. A magnetic field (corresponding to the Zeeman energy $\Delta_Z$) is applied along the NW.
The tight-binding Hamiltonian describing the NW has the form~\cite{Rainis2013}
	\begin{align}
	\begin{split}
	H &=\; \sum\limits_{j,s', s}
	c^\dag_{s',j+1}\left[-t\delta_{s' s} -\frac{i}{2a}\alpha_R\sigma^y_{s's} \right] c_{s,j} + H.c.\\
	&+\;\sum\limits_{j,s,'s} c^\dag_{s',j}\left[2t\delta_{s's} - \mu(x_j) \delta_{s's} + \Delta_Z\sigma^x_{s's}  \right] c_{s,j} 
	\\&+\;\sum\limits_{j} 	\Delta_{sc} \left(c^\dag_{\uparrow, j}
 c^\dag_{\downarrow,j} + H.c.\right), 
	\end{split}
	\label{eqn:Hamiltonian}
	\end{align}	
where $t=\hbar^2/(2m_0a^2)$ is the hopping amplitude, $c_{s,j}$ annihilates an electron with spin $s$ at site $j$ with coordinate $x_j$, $\mu(x)$  is a non-uniform chemical potential (measured from $E_{so}$), and $\sigma^{x,y}$ are the Pauli matrices. 
 
In the following we assume that the Zeeman energy is always larger than the superconducting gap, $\Delta_Z > \Delta_{sc}$. 
The chemical potential $\mu(x)$ is controlled by local gates and the NW is divided into a left and right section with lengths $L_-$ and $L_+$, respectively (see~Fig.~\ref{fig:setup}).
We model the non-uniformity of  $\mu(x)$ by a function with linear slope within the interface between the sections, $|x|<l_0$, where $l_0$ is the transition length,
\begin{align}
\mu(x) = \mu_{+}\Theta(x-l_0) + \mu_{+}\Theta(l_0-|x|)\frac{x-l_0}{2l_0}.
\label{eqn:chemical-potential}
\end{align}
Here, $\Theta(x)$ is the Heaviside step-function and $\mu_{+}$ is the value of  chemical potential in the right section.
In the left section, the chemical potential is set to the SOI energy, $\mu=0$, such that the left section is always in the topological phase. The gap in the left section 
is $\Delta_-\approx 2\Delta_{sc}\sqrt{E_{so}/\Delta_Z}$ for weak SOI, $E_{so} \ll \Delta_Z$, while $\Delta_-\approx \min\{\Delta_{sc}, |\Delta_Z-\Delta_{sc}| \}$ for strong SOI, $E_{so} \gg \Delta_Z$. The  gap in the right section 
is given by $\Delta_+ = |\Delta_Z - \sqrt{\Delta_{sc}^2 + \mu_+^2}|$ near the topological phase transition, $\mu_+\approx \mu^* = \sqrt{\Delta_Z^2 - \Delta_{sc}^2}$. The corresponding decay lengths 
are defined as $\xi_\pm = \hbar v_{F\pm}/\Delta_\pm$, where $v_{F\pm}$ are the 
Fermi velocities, depending on $\Delta_Z$ and $\mu_+$.

\begin{figure}
	\includegraphics[width=\columnwidth]{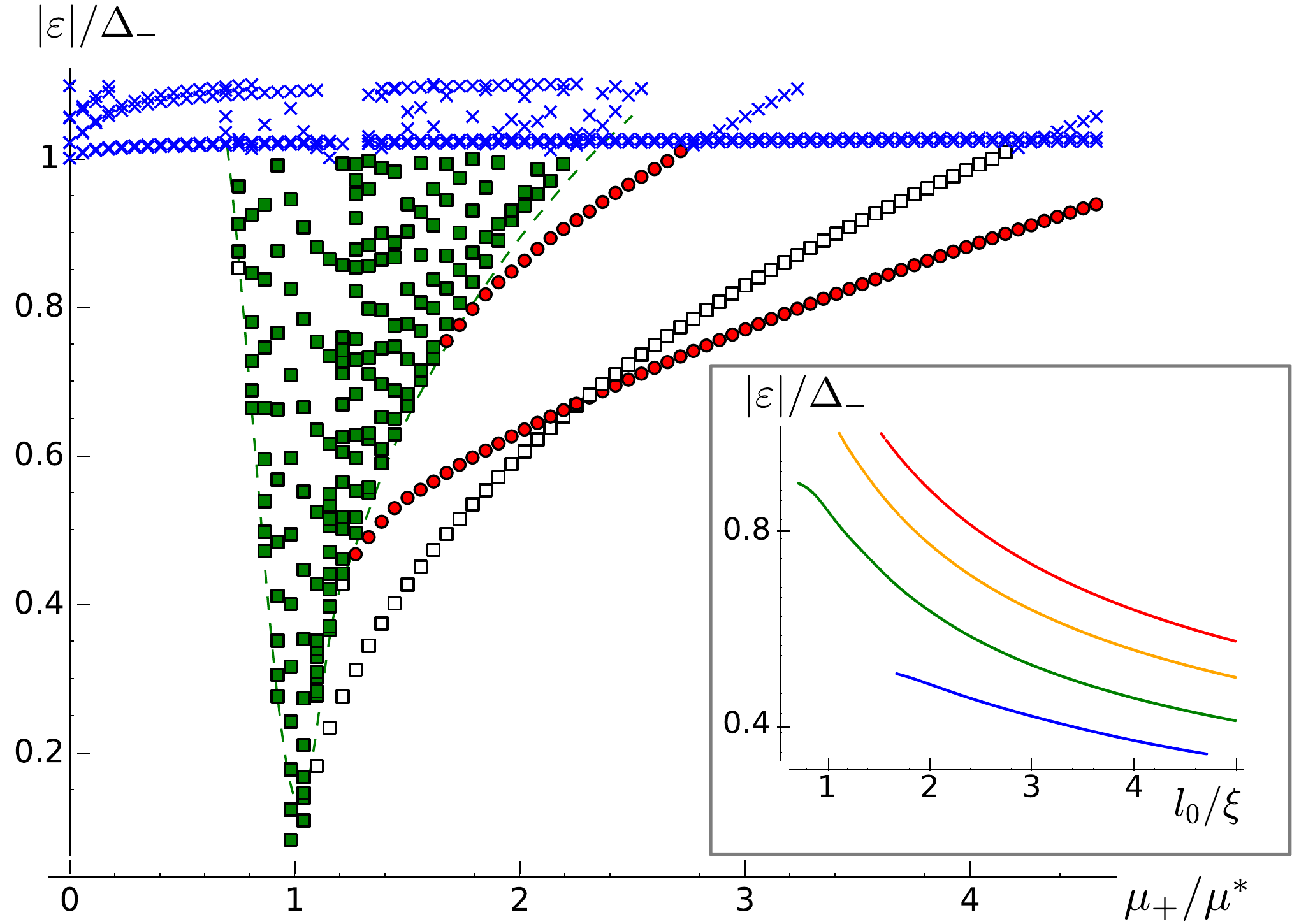}
\caption{Spectrum of the NW Hamiltonian Eq.~(\ref{eqn:Hamiltonian}) as function of $\mu_+$. The topological phase transition in the right NW section occurs at $\mu_+ = \mu^*$.  In addition to the bulk modes above the gap $\Delta_-$ (blue crosses), there are bulk modes in the right section (green squares) with energies above the gap $\Delta_+$ (green dashed line).
There also emerge subgap bound states: RABSs (black empty squares) and IABSs (red circles). The IABSs occur if $l_0\gtrsim \min\{\xi_+,\xi_-\}$
and have energies far below the gap $\Delta_-$ as $\mu_{+}$ approaches $\mu^*$.  The parameters chosen are $(E_{so}, \Delta_{sc}, \Delta_Z) = (0.05, 0.5, 1)\; \mathrm{meV}$,
corresponding to $(t, \Delta_-,\mu^*)=(10, 0.21, 0.87)\; \mathrm{meV}$ and $\xi_-\approx 30a$.
 For these parameters, $l_0 = 60a \approx 2\xi_-$ and  $L_{-} = L_{+} = 300a \approx 10\xi_-$. The inset shows the IABS energy as function of $l_0$ for different $\mu_+$: from bottom to top $\mu_+ /\mu^*=(1.3, 2, 3, 4)$.
The IABS energy decreases with increasing $l_0$ as $\varepsilon \propto l_0^{-1/2}$ \cite{SM}.}
\label{fig:modes}
\end{figure}

\mysection{Andreev bound states}
For $\mu_+ > \mu^*$~\cite{footnote1}, the right section is trivial, and one of the MBSs is located at the interface between the sections. 
 There also emerge a number of fermionic states with energies below the gaps $\Delta_\pm$ (see Fig.~\ref{fig:modes}): (i) Right Andreev bound states (RABSs) localized at the right end of the NW; (ii) Interface Andreev bound states (IABSs) localized at the interface between the sections~\cite{Chevallier2012, Kells2012, Liu2012, Bagrets2012,San-Jose2013,Roy2013,DeGottardi2013,DeGottardi2013a,Stanescu2014,Adagideli2014,Klinovaja2015, Fleckenstein2017}.
  The IABSs emerge when 
  $l_0$ exceeds the minimum of the two decay lengths, $l_0\gtrsim \min\{\xi_+,\xi_-\}$.
  The energies of IABSs can be significantly less than the gaps $\Delta_\pm$, if $\mu_{+}$ is close to the critical value, $\mu_+ \gtrsim \mu^*$, or if $l_0$ is much longer than $\xi_-$~(see Fig.~\ref{fig:modes}). In the latter case the energy of IABSs decreases with increasing $l_0$ as  $\varepsilon\sim \sqrt{\alpha_R \mu^* \mu_+/(2\Delta_Z l_0)}\propto l_0^{-1/2}$~\cite{SM} 
If there is a mechanism which allows to change the system from the GSs to some excited states, these IABSs will play a crucial role in determining the lifetime of the GS. 

\mysection{Lifetime of  GS in topological NW}
In physical setups the local electrostatic potentials are tuned by gates. The charges in the gates can fluctuate giving rise to fluctuations of the gate potentials. For simplicity, we consider a single normal metal gate located at distance $d_0$ from the NW. We denote the position of electrons in the gate by $\mathbf{R} = (R_x, \mathbf{R}_\bot)$, where $R_x$ is the coordinate along the NW, and $\mathbf{R}_\bot$ denotes the remaining coordinates. 
The gates are modeled by non-interacting electrons 
in $D$ dimensions (we consider $D=2,3$) and described by the gate Hamiltonian,
\begin{align}
H_g = \sum\limits_s\int d\mathbf{R}\; \Psi_s^\dag(\mathbf{R})\left[ -\frac{\nabla_{\mathbf{R}}^2}{2m_g} - \varepsilon_F^{(g)} \right] \Psi_s(\mathbf{R}),
\end{align}
where $\Psi_s$ is a field operator for electrons with spin $s$, $\varepsilon_F^{(g)}$ the corresponding Fermi energy, and $m_g$  the effective electron mass.  The electrostatic potential $\varphi(x_i)$ induced   by the gate electrons  at   site $i$ of the NW is given by
\begin{align}
\varphi(x_i) = e \sum\limits_s \int d\mathbf{R}\; \Psi_s^\dag(\mathbf{R})U(\mathbf{R}, x_i)\Psi_s(\mathbf{R}) ,
\end{align} 
where $e$ is electron charge and $U(\mathbf{R}, x_i)$  the potential at site $i$ of the NW created by a unit point charge located at position $\mathbf{R}$ in the gate.
For  screened electron-electron interactions it is given by the Yukawa potential~\cite{Kittel1986},
\begin{align}
U(\mathbf{R}, x_i) = \frac{e^{-d(\mathbf{R}, x_i)/\lambda}}{4\pi \epsilon \epsilon_0 d(\mathbf{R}, x_i)} .
\label{eqn:Yukawa}
\end{align}
Here,  $\epsilon$ is the dielectric constant of the NW, $\epsilon_0$ is the vacuum permittivity, $d(\mathbf{R}, x_i)$  the distance between the point $\mathbf{R}$ in the gate and the $i$-th site in the NW, and $\lambda \sim d_0$  the screening length. The main contribution to the potential created by the gate stems from electrons located in a layer of  width $\lambda$ near the gate surface, and we replace $U$ with a simplified delta-function potential, $U = U_0 \lambda^D \delta(x_i - R_x)\delta(\mathbf{R}_\bot)$, 
with amplitude  $U_0 \approx  e^{-d_0/\lambda}/(4\pi \epsilon \epsilon_0 d_0)$. We note that the formalism developed below remains valid for  general  $U$.
Finally, the coupling of electrons in the NW to the fluctuating electrostatic potential $\varphi$ is described by the  Hamiltonian
\begin{align}
H_{\varphi} = e\sum\limits_{j,s} \varphi(x_j) c^\dagger_{s,j}  c_{s,j}.
\end{align}
We focus on the gate fluctuations, $\delta \varphi = \varphi - \langle \varphi \rangle$, absorbing the  mean value $\langle \varphi \rangle$ in $\mu$.
\begin{figure}[t]
\includegraphics[width=\columnwidth]{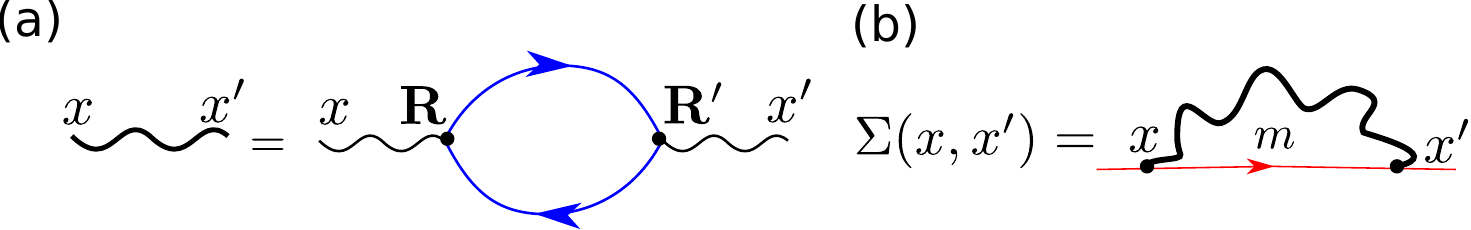}
\caption{
	(a) Feynman diagram corresponding to the correlators $\mathcal{D}^{R,A,K}$ (thick curly line) for fluctuations of the gate potential. The electrons in the NW interact with electrons in the gates with Green functions $\mathcal{G}_g^{R,K,A}$ (blue lines) via the  Coloumb potential $U(\mathbf{R}, x)$ (thin curly lines).
	(b) Diagram for the self-energy $\Sigma$ corresponding to the process where  gate fluctuations (thin curly line [see (a)]) promote the system from its GS to an excited state $m$ (straight red  line).
}
\label{fig:Feynman}
\end{figure}

We calculate the lifetime of the GS 
using Keldysh techniques~\cite{Keldysh1965diagram, KamenevLevchenkoAdvPhys2010}. The retarded Green function $\mathcal{D}^{R}$ for the bosonic fields $\delta \varphi$ corresponding to the Feynman diagram shown in Fig.~\ref{fig:Feynman}a is given by 
\begin{align}
\begin{split}
\mathcal{D}^R(x,x',t) &=\; i\int d\mathbf{R}d\mathbf{R'}\; U(\mathbf{R},x) U(\mathbf{R'},x')\\&\times\left[ \mathcal{G}^R_{g}(\mathbf{R}, \mathbf{R}',t) \mathcal{G}^K_{g}(\mathbf{R}',\mathbf{R},-t)
%\right.&\left.
\right.\\&\left.+\; 
\mathcal{G}^K_{g}(\mathbf{R},\mathbf{R}',t) \mathcal{G}^A_{g}(\mathbf{R}',\mathbf{R},-t)\right],
\end{split}
\label{eqn:retarded}
\end{align}
where, in energy representation, $\mathcal{G}^{R(A)}_g(\varepsilon)$ is the retarded (advanced) Green function for the gate electrons, $\mathcal{G}^{K}_g(\varepsilon)$ the Keldysh counterpart, and $\varepsilon$ is measured from the Fermi level $\varepsilon_F^{(g)}$.
\begin{figure*}[t]
	\includegraphics[width=\linewidth]{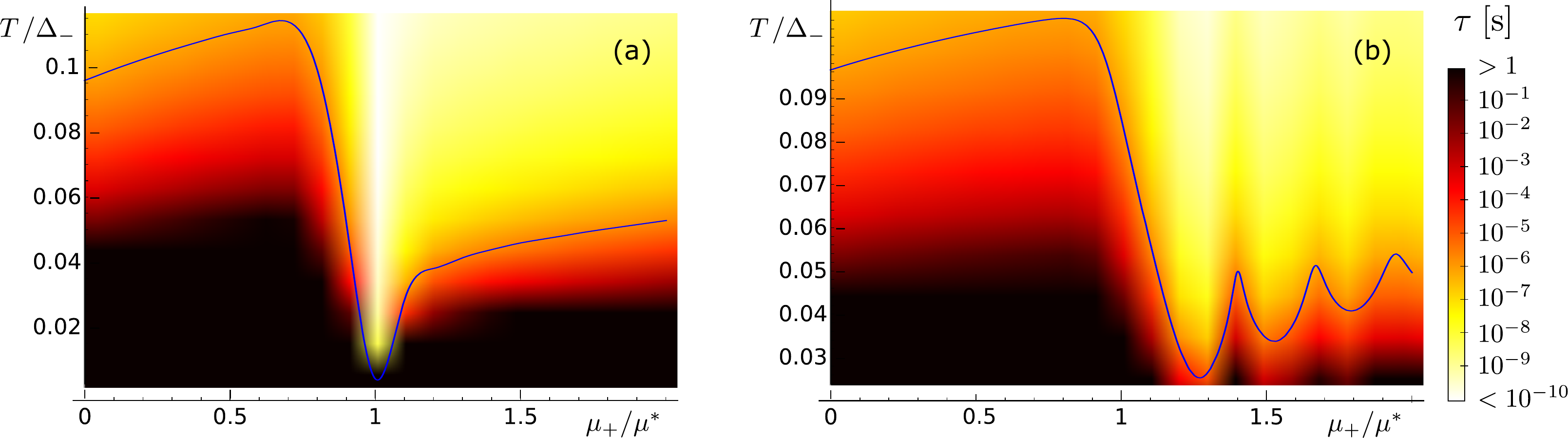}
	\caption{ Lifetime $\tau$  as  function of temperature $T$ and chemical potential $\mu_{+}$  for (a)  long  ($L_+ = 300a \approx 10\xi_-$) and (b)  short ($L_+ = 60a \approx 2\xi_-$) right NW section.  The blue lines hold for	
	$\tau(\mu_+, T_{\mu s}) = 1\;\mathrm{\mu s}$, and $\mu_{+} = 0$ corresponds to uniform $\mu(x)$.
	As $\mu_{+} \to \mu^*$, $\tau$ decreases, being much shorter for  case (a).
	If the right section is tuned to the trivial phase, $\mu_{+} \gg \mu^*$, the temperature required for maintaining the same $\tau$ becomes approximately twice smaller than for uniform $\mu(x)=\mu_{+}=0$. While for $\mu_{+} < \mu^*$,
	  the required  $T$ is similar for (a) and (b),  $\tau$ and $T$ oscillate for $\mu_+> \mu^*$ for (b) but not for (a) \cite{SM}.	 		
		  The parameters chosen are $L_{-} = 300a = 10\xi_-$, $\Delta_-=2.4\;\mathrm{K}$, $\varepsilon_F^{(g)} = 10\;\mathrm{eV}$, $v_F^{(g)} = 2\cdot 10^6\;\mathrm{m/s}$, $d_0 = 100\;\mathrm{nm}$, $\lambda = \mathrm{300}\;\mathrm{nm}$, $\epsilon = 15$, and the rest as in Fig.~\ref{fig:modes}.
		 The value for the effective coupling constant [see~Eq.(\ref{eqn:DR})] is then given by $g=1.05$.
}
	\label{fig:decoh-long}
\end{figure*}

\begin{figure}[b]
	\includegraphics[width=\linewidth]{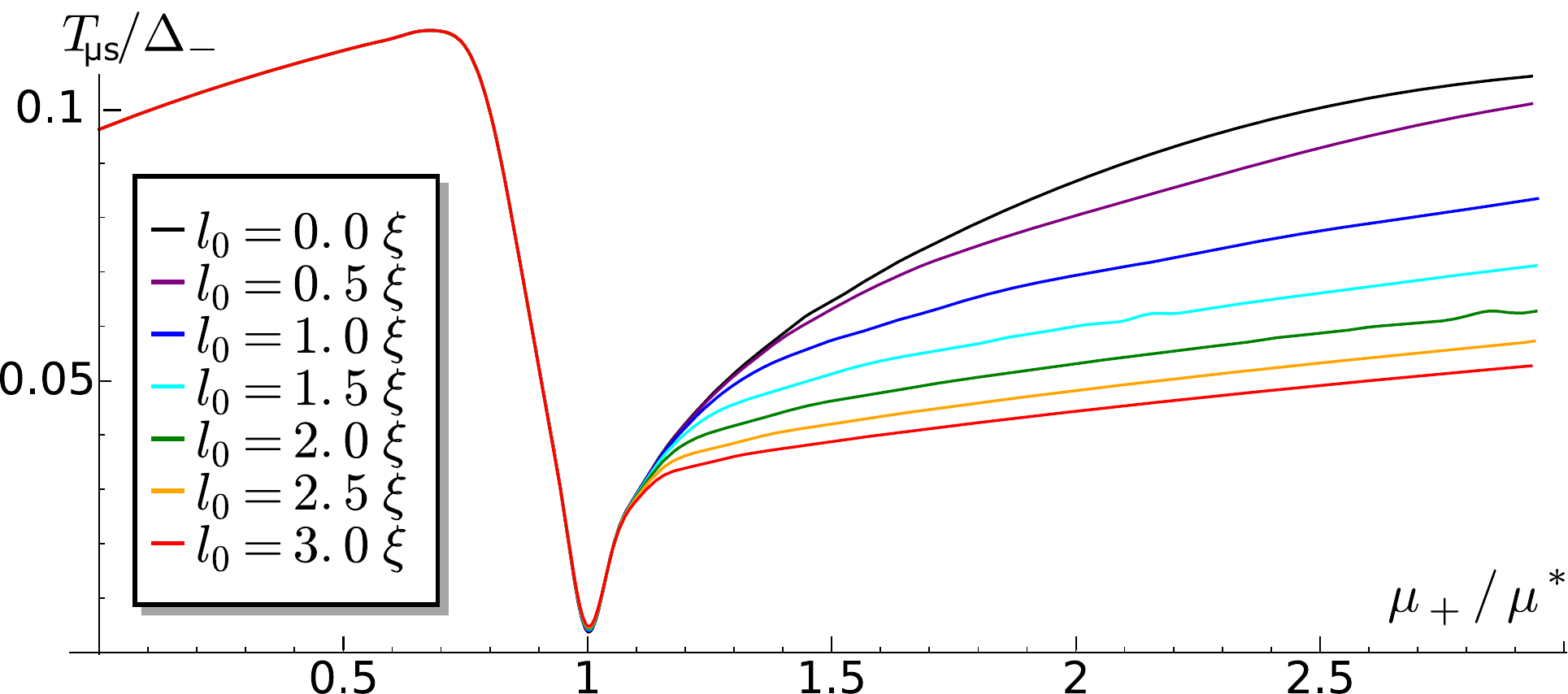}
	\caption{	
		Temperature $T_{\mu s}$ for which $\tau \gtrsim 1\;\mathrm{\mu s}$ as function of $\mu_+$ plotted for different values of 	
		$l_0$. While the dip at $\mu_{+}\approx \mu^*$ does not depend on $l_0$, for $\mu_{+}> \mu^*$, $T_{\mu s}$ is determined by the  IABS energy which decreases with increasing $l_0$. For an abrupt transition ($l_0=0$, black top line), $T_{\mu s}$ for a trivial right section ($\mu_{+} \gg \mu^*$)
		remains practically the same as for a uniform chemical potential, $\mu(x)=\mu_{+} = 0$. The parameters are the same as for Fig.~\ref{fig:decoh-long}.}
	\label{fig:decoh-long-l0}
\end{figure}

For a 2D or 3D gate, the correlations  are short-ranged and we find $\mathcal{G}^R_g(\mathbf{R}, \mathbf{R}', 0) \approx - i \pi  \nu_D k_F^{-D} \delta(\mathbf{R}-\mathbf{R}')$~\cite{KamenevLevchenkoAdvPhys2010}, where $\nu_D$ is the $D$-dimensional density of states at the Fermi level and $k_F$ the Fermi wavevector in the gate. 
For $\mathcal{G}_g^K$ we use the fluctuation-dissipation theorem (FDT), $\mathcal{G}_g^K (\varepsilon)= 2i \tanh(\varepsilon/2T)\mathrm{Im}\mathcal{G}^R(\varepsilon)$. Performing the integrations in Eq.~(\ref{eqn:retarded})  we then get 
\begin{align}
\mathcal{D}^R(x,x',\omega) = -ig \omega \delta(x-x'),
\label{eqn:DR}
\end{align}
where  $g = \pi  (\nu_DeU_0)^2 \lambda^{D}/k_F^D$ is the effective coupling constant. The Keldysh version again follows from the FDT, 
$\mathcal{D}^K (\omega)= 2i \coth(\omega/2T) \mathrm{Im} \mathcal{D}^R(\omega)$.

Importantly, the fluctuations $\delta \varphi$ can excite the non-local fermion (shared by two MBSs) out of the GS. The Feynman diagram for the Keldysh self-energy $\Sigma$ corresponding to this process is shown in Fig.~\ref{fig:Feynman}b. The real part of the retarded self-energy integrated over coordinates, $\Sigma^R(\varepsilon) = \int dxdx'\; \Sigma^R(x,x',\varepsilon)$, corresponds to a shift of the energies of the MBSs and vanishes if one disregards the overlap between MBSs. The imaginary part $\Gamma = \mathrm{Im}\; \Sigma^R(\varepsilon=0)$ equals to the decay rate of the process shown in Fig.~\ref{fig:Feynman}b and is related to the lifetime $\tau = \hbar/ \Gamma$ the system stays in the degenerate GS before being excited by fluctuations $\delta \varphi$ out of the GS into some excited states. In leading order in $g$ we get
\begin{multline}
\Sigma^R(\varepsilon) = \sum\limits_m\int dx dx' \rho_{Rm}(x)\left[G^R_m(\varepsilon - \omega) \mathcal{D}^K(x,x',\omega)\right.
\\\left.+G^K_m(\varepsilon - \omega)\mathcal{D}^A(x,x',\omega)
\right] \rho_{Rm}^*(x') \frac{d\omega}{2\pi},
\end{multline}
where $\rho_{Rm}(x) = \bar{\Phi}_R(x) \tau_z \Phi_m(x)$, and $\Phi_m$ ($\Phi_R$) is the eigenspinor of  $H$ [see  Eq.~(\ref{eqn:Hamiltonian})] written in the basis $(c_{\uparrow j},  c_{\downarrow j}, c_{\downarrow j}^\dag,  -c_{\uparrow j}^\dag)$  corresponding to the $m$-th excited state (right MBS) with energy $\varepsilon_m\neq 0 $.
The Pauli matrix $\tau^z$ acts in the particle-hole space, and $G^R_m = 1/(\varepsilon - \varepsilon_m + i0)$ and $G^K_m = -2\pi i \delta(\varepsilon-\varepsilon_m)\tanh(\varepsilon_m/2T)$ are retarded and Keldysh Green functions for the NW electrons, respectively. Finally, we obtain the lifetime as
\begin{align}
\tau^{-1} = \frac{g}{\hbar}\sum\limits_m \frac{\varepsilon_m}{\sinh \left(\varepsilon_m/T \right)} \int dx\;  \left|\rho_{Rm}(x)\right|^2.
\end{align}
In Fig.~\ref{fig:decoh-long} we plot $\tau$ as function of $\mu_{+}$ and $T$ and estimate the temperature $T_{\mu s}$ required to maintain the lifetime of order  $1\;\mathrm{\mu}s$, {\it i.e.}, $\tau(\mu_+, T_{\mu s}) = 1\;\mathrm{\mu s}$. For uniform chemical potentials, $\mu_+ = 0$, the results agree with Ref.~\onlinecite{Schmidt2012}. For non-uniform $\mu(x)$,  $\tau$ reaches its minimum if $\mu_{+}\approx \mu^*$. If the right section is long so that  $L_+/\xi_- \gtrsim 10$ (see Fig.~\ref{fig:decoh-long}a), $T_{\mu s}$ must be as low as $\Delta_-/200$ at $\mu_+ = \mu^*$ and in the range from $\Delta_-/20$ to $\Delta_-/10$ for
larger $\mu_+$. If $l_0$ increases, the IABS energy decreases, and even lower $T_{\mu s}$'s are required at $\mu> \mu^*$ (see Fig.~\ref{fig:decoh-long-l0}). Thus, if one takes $\Delta_-\sim 0.1-0.2\;\mathrm{meV}$~\cite{DeMoor2018}, $T_{\mu s}$ can 
be in the range $1-100\;\mathrm{mK}$ depending on $\mu_+$ and $l_0$. 
In order to maintain $\tau\sim 1\;\mathrm{ms}$ one has to reduce the temperature at least by a factor of two.
These estimates are relevant for Majorana qubit proposals as described in Ref.~\onlinecite{Alicea2011}.  

For short right sections, $L_{+} \sim l_0 \sim \xi_-$ (see Fig.~\ref{fig:decoh-long}b), $T_{\mu s}$ must be as low as $\Delta_-/50$, if $\mu_+$ is close to $\mu^*$. If $\mu_+$ is away from $\mu^*$,  $T_{\mu s} \sim \Delta_-/30 - \Delta_-/20$. Also, $\tau$ and $T_{\mu s}$ oscillate with $\mu_+$, so that a slight change in $\mu_+$ can lead to large variations of $\tau$ by several orders of magnitude, and fine-tuning of $\mu_+$ is required to maintain longer lifetimes. This situation does not occur in the qubit proposal of Ref.~\onlinecite{Alicea2011}, since the trivial section located between two topological sections has to be significantly longer than $\xi_+$ \cite{Zyuzin}. However, it can be relevant in schemes where a gate located near the end of the NW is used to couple MBSs to a quantum dot~\cite{Golub2011, Zocher2013, Gong2014, Leijnse2014, Vernek2014, Deng2016, Szombati2016, Hoffman2017, Chevallier2018}. If the gate tunes the ending of the NW out of the topological phase,  $\tau$ can decrease hundreds or even thousands of times. If the right section is in the normal state with a discrete spectrum  (similar to a quantum dot), the oscillations of $T_{\mu s}$ and $\tau$ are even more pronounced, and $T_{\mu s}$ can be ten times lower (see SM \cite{SM}).

We note that if one has to bring a part of the NW into the topological (trivial) phase during manipulations of MBSs or reading out a qubit, the drastical shortening of $\tau$ at $\mu\approx\mu^*$ can be problematic. On one hand all manipulations via gates must be adiabatically slow. On the other hand, the time span during which the local chemical potential is close to $\mu^*$ should not  exceed the bottleneck lifetime $\tau(\mu_+=\mu^*, T)$. And even if $\mu_+$ is significantly bigger than $\mu^*$, 
$\tau$ for a non-uniform $\mu(x)$ can be orders of magnitude shorter than in the uniform case, $\mu_+ = 0$~\cite{Schmidt2012}.
However, our results show that one can achieve longer lifetimes (or maintain the same $\tau$ at higher temperatures) by a better screening of the gates making the transition between the two sections more abrupt. Another option could be to increase  the bulk gaps $\Delta_{\pm}$ by taking a material with larger $E_{so}$ or by tuning the tunnel-coupling to the superconductor and changing the proximity gap $\Delta_{sc}$  \cite{reeg_1,reeg_2,lutchyn,das_sarma,flensberg,reeg_3, DeMoor2018}. 
One can also optimize the gate designs and reduce the coupling constant $g$. 

\mysection{Conclusions}
We have studied the lifetime $\tau$ of the GS due to charge fluctuations on a
nearby gate for a simple model of a topological NW with a non-uniform chemical potential. If the NW is divided into topological and trivial sections, there emerge IABSs with energies significantly below the gap. The noisy gates can excite the system from its GS to one of the IABSs reducing the lifetime of the GS. If one braids MBSs by changing gate potentials~\cite{Alicea2011}, the lifetime becomes extremely short if $\mu_+$ is close to a critical value $\mu^*$. Even if $\mu_+ > \mu^*$ our estimations show (see Fig.~\ref{fig:decoh-long}) that $\tau$ is hundreds or even thousands times shorter than for uniform chemical potentials. In order to maintain $\tau \sim \mu s$ one has to keep the temperature below $50\;\mathrm{mK}$ if the topological gap is of order of $0.1\;\mathrm{meV}$. This estimation can also be used for short right sections, however, a fine-tuning of $\mu_+$ is required to achieve longer $\tau$'s. The mechanism restricting the lifetime of MBSs considered here can be dominant in comparison to the one related to the quasiparticle poisoning  \cite{Rainis2012} in case of a high tunneling resistance of a barrier between the NW and the superconductor. However, if the length of the left section is of order of the correlation length $L_- \gtrsim \xi_-$, different mechanisms of the decoherence due to coupling of overlapping MBSs to electromagnetic environment or to phonons~\cite{Knapp2018} may be more relevant.

This work was supported by the Swiss National Science Foundation (Switzerland) and by the NCCR QSIT. This project received funding from the European Union’s Horizon 2020 research and innovation program (ERC Starting Grant, grant agreement  No 757725).

\widetext
\begin{center} 
\textbf{Supplemental Material for\\Lifetime of Majorana qubits in Rashba nanowires with  non-uniform chemical potential} 
\end{center} 

\section{A. Interface Andreev bound states} 
In order to estimate the energy of the IABS localized at the interface between topological and trivial phases  analytically, we replace the tight-binding Hamiltonian,~Eq.~(1), with the following continuum Hamiltonian, describing the long-wavelength and low-energy physics $|\varepsilon|\ll t$,
\begin{align}
H &=\; \int dx\; \bigg( \sum\limits_{s} 
c^\dag_{s}(x)\left[-\frac{\partial_x^2}{2m_0} - \mu\right]c^\dag_{s}(x) + \sum\limits_{s,s'} c^\dag_{s'}(x)\left[-i\alpha_R\sigma^y_{s's}\partial_x + \Delta_Z\sigma^x_{s's}\right] c_{s}(x) 
+	\Delta_{sc} \left(c^\dag_{\uparrow}(x)
c^\dag_{\downarrow}(x) + H.c.\right)\bigg).
\label{eqn:Hamiltonian}
\end{align}
For a uniform $\mu(x)=\mu$ the gap at zero momentum is given by $\Delta_\mu = |\Delta_Z^2 - \sqrt{\Delta_{sc}^2 + \mu^2}|$, and near the topological phase transition, $\mu\approx \mu^*$, it can be estimated as $\Delta_\mu\approx\mu^*|\mu-\mu^*|/\Delta_Z$.

For a smoothly varying chemical potential $\mu(x)$ given by Eq.~(2) with $l_0\gg \xi_-$, we treat the Hamiltonian,~Eq.~(\ref{eqn:Hamiltonian}), quasiclassically using Wentzel-Kramers-Brillouin~(WKB) approximation~\cite{LandauLifshitz3}. We calculate the quasiclassical momentum $p(x)$ near the point $x^*$ in the transition region such that $\mu(x^*) = \mu^*$. Since the gradient of $\mu(x)$ equals $\mu_+/(2l_0)$, the quasiclassical momentum can be estimated as
\begin{align}
p(\varepsilon,x) = \frac{\sqrt{\varepsilon^2 - \Delta_{\mu(x)}^2}}{\alpha_R} =\frac{\sqrt{\varepsilon^2 - \left[|x-x^*|\mu^* \mu_+/(2\Delta_Z l_0)\right]^2}}{\alpha_R}.
\end{align}
The quasiclassical momentum equals zero at the turning points $x^*\pm x_0$, where $x_0 = 2l_0 \Delta_Z \varepsilon/(\mu^* \mu_+)$, and the Bohr--Sommerfeld condition for the energy of the bound state reads
\begin{align}
\int\limits_{x^*-x_0}^{x^*+x_0} p(\varepsilon, x) dx = \frac{\pi}{2}.
\end{align}
Performing the integration we obtain the following estimation for the energy of IABS:
\begin{align}
\varepsilon \sim \sqrt{\frac{\alpha_R \mu^* \mu_+}{2\Delta_Z l_0}},
\end{align}
which is in agreement with the results obtained numerically (see inset of Fig.~2 in the main text). Unlike the conventional long SNS junctions where the energy of the lowest Andreev bound state is inverse proportional to the length of the normal section~\cite{Ishii1970,Kulik1970,Bardeen1972, Dmytruk2018}, the energy of IABSs decays with increasing length $l_0$ as $\varepsilon\propto l_0^{-1/2}$.

\section{B. Short right section}

\subsection{Superconducting right section}
\begin{figure}
	\includegraphics[width=\linewidth]{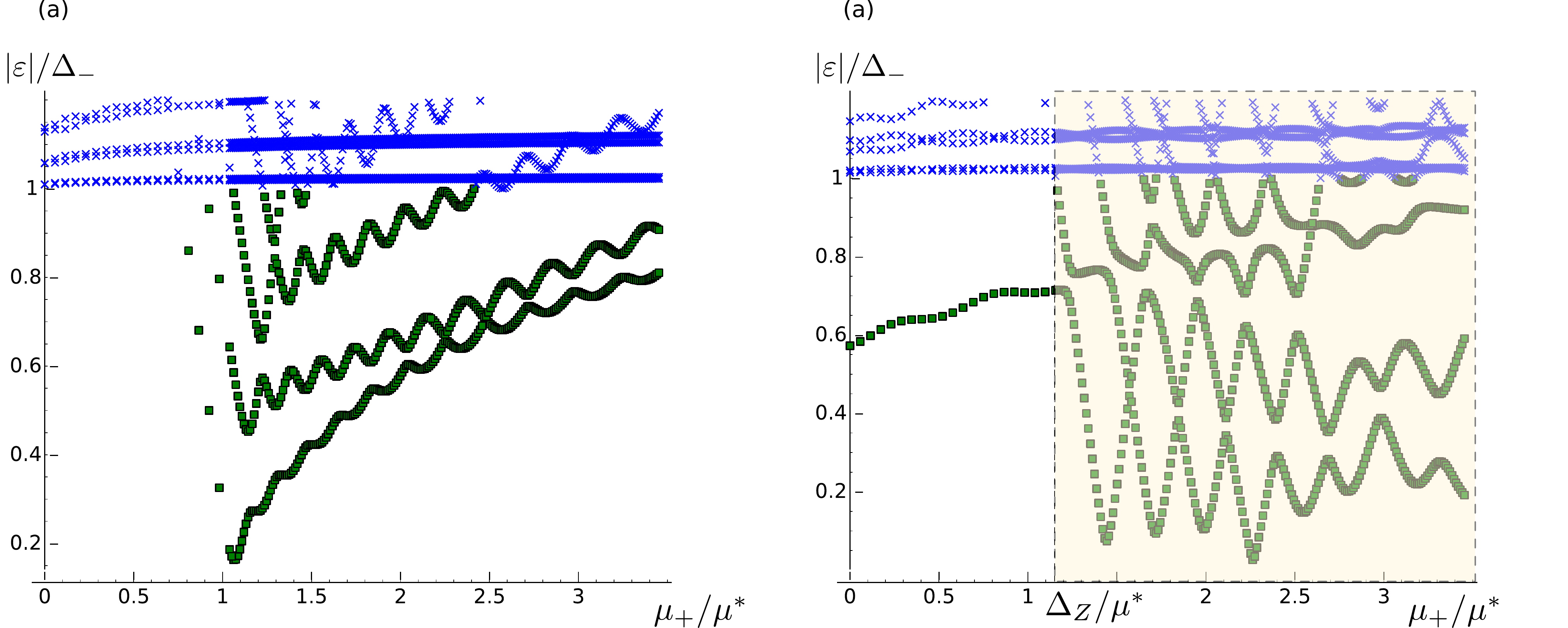}
	\caption{(a) Spectrum of the NW [obtained numerically by diagonalizing Eq.~(1) of the main text] as function of $\mu_+$ for a short right section $L_+ = l_0 = 2\xi_-$ and a uniform $\Delta_{sc}$ over the entire NW length. If $\mu_+$ is larger than the critical value, $\mu_+ > \mu^*$, in addition to the bulk modes above the gap $\Delta_-$ (blue crosses), there emerge bulk modes in the right section (green squares), whose energies show oscillatory dependence on $\mu_+$ approaching  the gap $\Delta_-$ for $\mu_+ \gg \mu^*$. The parameters used are the same as for Fig.~2 of the main text.
(b) The same as (a) but for a normal right section ($\Delta_{sc}=0$). The spectrum of a normal section can be only partially gapped by the magnetic field.
% and consists of exterior and interior branches.
If $\mu_+ < \Delta_Z$, the spectrum of the NW Hamiltonian, see Eq.~(1) of the main text, consists of states above the gap $\Delta_-$ localized in the left superconducting section (blue crosses) and states  localized in the right normal section (green squares) originating from the exterior branches of the spectrum with quantized energies, whereas the interior branches of the spectrum in the right section are gapped.  If $\mu_+$ becomes larger than $\Delta_Z$ (shaded region) the number of states with energies below $\Delta_-$ increases since both exterior and interior branches of the spectrum are ungapped at these energies. The energy levels exhibit characteristic oscillations as $\mu_+$ is changed.}
	\label{fig:spectra}
\end{figure}

\begin{figure}[b]
	\includegraphics[width=0.5\linewidth]{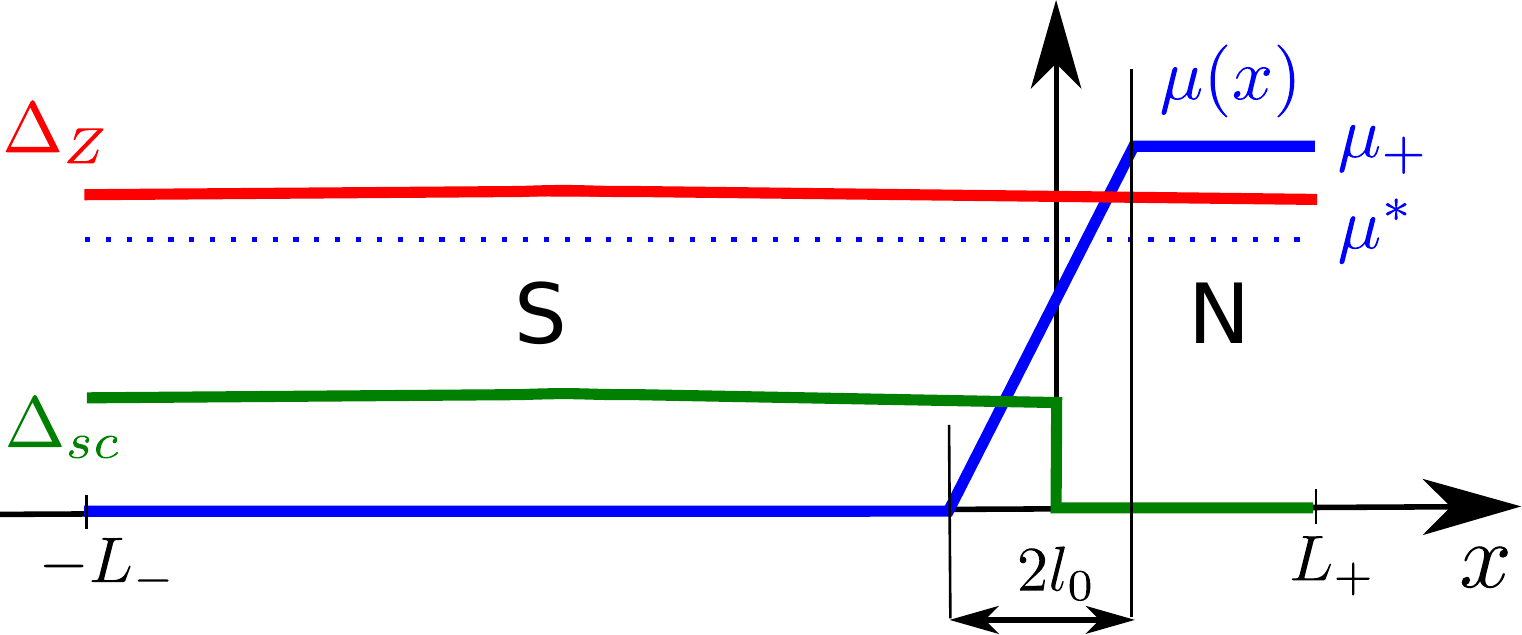}
	\caption{Spatial dependence of the parameters in an SN junction. The non-uniform chemical potential $\mu(x)$  is controlled by a nearby gate similar to the setup shown in Fig.~1 of the main text. The proximity-induced superconducting gap is non-zero, $\Delta_{sc}\neq 0$, in the left (S) section, and vanishes in the right (N) section. The Zeeman energy $\Delta_Z$ is uniform.}
	\label{fig:setup-normal}
\end{figure}

First, we consider a setup in which both the left and the right sections are superconducting, and the right section is short $L_+\sim\xi_-$ (see Fig.~1 of the main text). In this case, we also assume that $\Delta_{sc}$ is uniform. The spectrum is obtained by diagonalizing the tight-binding Hamiltonian defined by Eq. (1) of the main text and is  shown in Fig.~\ref{fig:spectra}a. If the chemical potential in the right section $\mu_+$ is below the critical value, $\mu_+ < \mu^*$, both the left and the right sections are in the topological phase, the MBSs are localized at the left end of the left section and at the right end of the right section, and they are well separated from the rest of the excitations by a superconducting gap $\Delta_-$.  If $\mu_+$ is above the critical value, $\mu_+>\mu^*$, there emerge IABSs in the right section with energies below $\Delta_-$. Since the right section is short $L_+ \sim \xi_-$, it is impossible to distinguish between the bulk states in the right section and the interface states. We note that in the vicinity of the critical value $\mu_+ \approx \mu^*$, the energy of such IABSs can be as low as $\hbar v_{F+}/L_+$. Due to the finite size of the right section, the energies oscillate as $\mu_+$ grows, which leads to oscillations of the lifetime $\tau$  (see Fig. 4b of the main text).

\subsection{Normal right section}
In order to consider the case where the topological NW is coupled to a quantum dot we consider also the case of an SN-junction, assuming that the right section is normal and $\Delta_{sc}$ vanishes in the right section. To account for a non-uniform proximity-induced superconducting gap $\Delta_{sc}(x)$, we slightly modify the tight-binding Hamiltonian defined in Eq.~(1) of the main text:
	\begin{align}
\begin{split}
H &=\; \sum\limits_{j,s', s}
c^\dag_{s',j+1}\left[-t\delta_{s' s} -\frac{i}{2a}\alpha_R\sigma^y_{s's} \right] c_{s,j} + H.c.
+\sum\limits_{j,s,'s} c^\dag_{s',j}\left[2t\delta_{s's} - \mu(x_j) \delta_{s's} + \Delta_Z\sigma^x_{s's}  \right] c_{s,j} 
\\&+\;\sum\limits_{j} 	\Delta_{sc}(x_j) \left(c^\dag_{\uparrow, j}
c^\dag_{\downarrow,j} + H.c.\right).
\end{split}
\label{eqn:tb-Hamiltonian}
\end{align}	
We model the non-uniformity of  $\Delta_{sc}(x)$ by a function with linear slope within the interface between the sections that vanishes in the right section:
\begin{align}
\Delta_{sc}(x) = \Delta_{sc}\left[\Theta(-l_\Delta - x) - \Theta(l_\Delta-|x|)\frac{x+l_\Delta}{2l_\Delta}\right],
\label{eqn:Delta-sc}
\end{align}
where $l_\Delta$ is a characteristic transition length for the proximity gap. However, our calculations show that the lifetime $\tau$ does not depend much on $l_\Delta$, and here we focus on the case of an abrupt drop of $\Delta_{sc}$ in the right section, $l_\Delta \ll \xi_-$, so that the spatial dependence of the proximity gap $\Delta_{sc}(x)$ is stepwise~(see Fig.~\ref{fig:setup-normal}), 
\begin{align}
\Delta_{sc}(x) = \Delta_{sc}\Theta(-x).
\label{eqn:Delta-sc-stepwise}
\end{align}

The spectrum for the normal part of the NW is partially gapped and consists of two branches: exterior and interior~\cite{Klinovaja2012}.
The spectrum of the Hamiltonian, see~Eq.~(\ref{eqn:tb-Hamiltonian}), describing the SN-junction is shown in Fig.~\ref{fig:spectra}b. If $\mu_+ < \Delta_Z$ only the exterior branch states have energies close to the chemical potential. If $\mu_+ > \Delta_Z$, also the interior branch states lies close to the chemical potential. The energies of localized states oscillate rapidly as $\mu_+$ changes. A qualitative description of these oscillations can be obtained if one disregards coupling between the exterior and interior branches at the boundaries. The energies of the states localized in the normal part satisfy a condition of the following general form:
\begin{align}
\left[\frac{\varepsilon_n}{\hbar v^{i(e)}_{F+}(\mu_+)} + k_F^{i(e)}(\mu_+)\right]L_+ \; \mathrm{mod}\; 2\pi = \chi^{i(e)}(\mu_+ + \varepsilon_n),
\end{align}
where $v^{i(e)}_{F,+}(\mu_+)$ and $k_F^{i(e)}(\mu_+)$ are Fermi velocity and Fermi wavevector of the interior~(exterior) branch, respectively, and the phase $\chi^{i(e)}(\mu_+ + \varepsilon_n)$ depends on the specific boundary conditions. While a derivation of an explicit form of  $\chi^{i(e)}(\mu_+ + \varepsilon_n)$ is complicated for the general case, one can get a simple estimate by disregarding the variation of $\chi^{i(e)}$ for small changes of $\mu_+$. In this case, the period of oscillations is of order of $h v_{F+}^{i(e)}/L_+$, and the oscillations are more rapid for $\mu_+>\Delta_Z$ when the interior modes emerge in the right section, since $v_{F+}^i < v_{F+}^e$.

\begin{figure}
	\includegraphics[width=\linewidth]{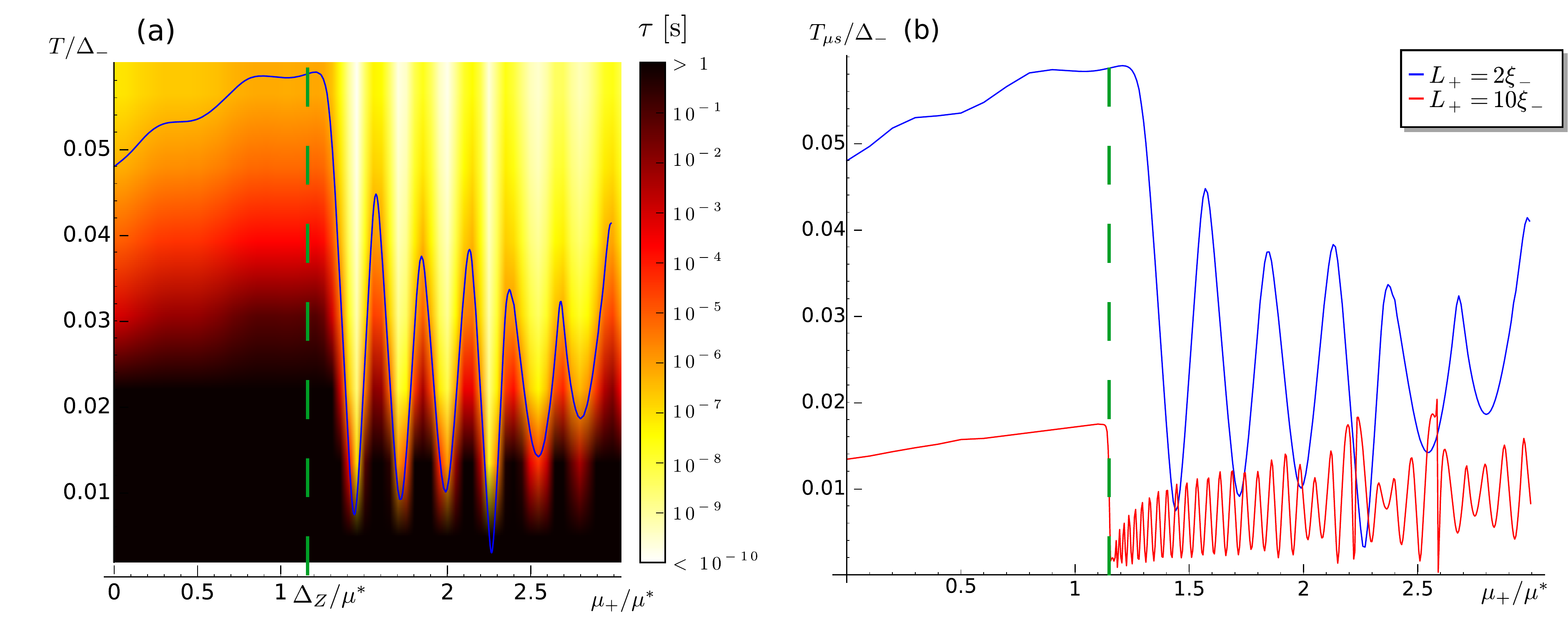}
	\caption{(a) Lifetime $\tau$  as  function of temperature $T$ and chemical potential $\mu_{+}$ for an SN junction with  short right (N) section, $L_+ = 30a = \xi_-$, see Fig.~\ref{fig:setup-normal}.  The blue line holds for	
		$\tau(\mu_+, T_{\mu s}) = 1\;\mathrm{\mu s}$. As $\mu_+$ becomes larger than $\Delta_Z$ (shown with a green dashed line), the lifetime $\tau$ and $T_{\mu s}$ oscillate with increasing $\mu_+$ with a period of order $hv_F^{(i)}/L_+$. The parameters chosen are the same as for Fig.~4.
	(b) Temperature $T_{\mu s}$ as function of $\mu_+$ plotted for different lengths of the right section $L_+$. For $\mu<\Delta_Z$ the temperature $T_{\mu s}$ decreases as $L_+$ increases, $T_{\mu s} \propto L_+^{-1}$. For larger values of the chemical potential $\mu_+ > \Delta_Z$, the oscillations of $T_{\mu s}$ become more rapid when the right section increases.}
	\label{fig:lifetime-NS-short}
\end{figure}

In Fig.~\ref{fig:lifetime-NS-short}a we plot the lifetime $\tau$ as function of $\mu_{+}$. The lifetime $\tau$ and the temperature $T_{\mu s}$ oscillate with $\mu_+$ for $\mu_+>\Delta_Z$, so that a slight change in $\mu_+$ can lead to large variations of $T_{\mu s}$. These oscillations are even more pronounced here than in case of the superconducting right section (see Fig. 4b), so that $T_{\mu s}$ can be in the range from $\Delta/200$ to $\Delta/25$, depending on tuning of $\mu_+$. In Fig.~\ref{fig:lifetime-NS-short}b  we compare how $T_{\mu s}$ depends on $\mu_+$ for different lengths $L_+$. Since the number of discrete levels in the normal section grows with increasing $L_+$, the temperature $T_{\mu s}$ decreases for longer right sections. For large values of the chemical potential $\mu_+ > \Delta_Z$, the oscillations of $T_{\mu s}$ become more rapid for increased right section lengths, the period being of order $h v_F^{(i)}/L_+$.
 
%\begin{figure} 
%	\includegraphics[width=0.5\linewidth]{density_matplotlib_NS_short.pdf}
%	\caption{Temperature $T_{\mu s}$ as function of $\mu_+$ plotted for different length of the transition region for a short normal right section, $L_+ = 60a = 2\xi_-$. The oscillations of $T_{\mu s}$ have a larger amplitude in case of shorter transition lengths.}
%	\label{fig:lifetime-vs-l0}
%\end{figure}

%\begin{figure}
%	\includegraphics[width=0.5\linewidth]{T_vs_dmu_NS_mod.pdf}
%	\caption{Temperature $T_{\mu s}$ as function of $\mu_+$ plotted for different length of the right section. For $\mu<\Delta_Z$ the temperature $T_{\mu s}$ decreases as the length of the right section becomes longer, $T_{\mu s} \propto L_+^{-1}$. For higher values of chemical potential $\mu_+ > \Delta_Z$, the oscillations of $T_{\mu s}$ are more rapid for the longer right section.}
%	\label{fig:lifetime-vs-Lplus}
%\end{figure}


\begin{thebibliography}{99}
\bibitem{footnote1}{
	We consider only $\mu_+ \geq 0$, which is a reasonable assumption for weak SOI, $E_{so} < \Delta_Z$, while for strong SOI the results are the same for $\mu_+ \leq 0 $, provided $\mu_+$ is not close to the bottom of the band.
}
\end{thebibliography}
\end{document}